\documentclass[dvipdfmx]{ws-ijmpb}
\usepackage{amsmath,bm,graphicx}
\usepackage{amsmath}
\usepackage{amssymb}
\usepackage{amsfonts}
\usepackage{graphicx,color}

\begin{document}

\markboth{T. Furuta {\it et al.}}
{Constitutive law for granular slow frictional drag}

%
\catchline{}{}{}{}{}
%

\title{Geometry-dependent constitutive law for granular slow frictional drag}

\author{T. Furuta}

\address{Department of Earth and Environmental Sciences, Nagoya University, Furocho, Chikusa, Nagoya 464-8601, Japan}

\author{K. Katou}

\address{Division of Biological Science, Graduate School of Science, Nagoya University, Furocho, Chikusa, Nagoya 464-8601, Japan}

\author{S. Itoh\footnote{Present address: Center for Gene Research, Nagoya University, Japan}}

\address{Division of Material Science (Physics), Graduate School of Science, Furocho, Chikusa, Nagoya 464-8601, Japan}

\author{K. Tachibana}

\address{Instrument Development Center of School of Sciences, Nagoya University, Furocho, Chikusa, Nagoya 464-8601, Japan}

\author{S. Ishikawa}

\address{Instrument Development Center of School of Sciences, Nagoya University, Furocho, Chikusa, Nagoya 464-8601, Japan}

\author{H. Katsuragi}

\address{Department of Earth and Environmental Sciences, Nagoya University, Furocho, Chikusa, Nagoya 464-8601, Japan\\
katsurag@eps.nagoya-u.ac.jp
}

\maketitle

\begin{history}
\received{Day Month Year}
\revised{Day Month Year}
\end{history}

\begin{abstract}
Frictional constitutive law for very slow vertical withdrawing of a thin rod from a granular bed is experimentally studied. Using a very precise creep meter, geometry-dependent granular frictional constitutive law is particularly examined. In some previous works, a dimensionless number $I=\dot{\gamma}D_g/\sqrt{p/\rho_g}$ has been used to characterize granular frictional constitutive laws, where $\dot{\gamma}$, $D_g$, $p$, and $\rho_g$ are the shear strain rate, grain diameter, confining pressure, and bulk density of granular bed, respectively. It has been considered that granular frictional constitutive law expressed by $I$ is universal (almost geometry-independent) in dense flow regime. In this study, however, we find that the geometry of the system is much more crucial to characterize granular friction in a very slow withdrawing regime. Specifically, the ratio between rod and grain diameters must be an essential parameter to describe the granular frictional constitutive law. Physical meaning of the geometry-dependent constitutive law is discussed on the basis of grains-contact-number dependence of granular behavior.
\end{abstract}


\section{Introduction}
Frictional sliding of granular matter can be found in a lot of natural phenomena. Examples include snow avalanche flow, seismic slip of a fault gouge layer, landsliding, and so on. Despite the ubiquity of granular friction in nature, its physical understanding has not yet been sufficient to properly control and/or predict the granular-related sliding states. One of the most famous systematic studies on granular friction has been reported by GDR Midi~\cite{GDRMiDi2004}. In the work, various experimental setups have been used to verify the universality of granular frictional constitutive law. 

To express the granular frictional property, a dimensionless number called inertial number, 
\begin{equation}
I=\frac{\dot{\gamma}D_g}{\sqrt{p / \rho_g}},
\label{eq:inertial_number}
\end{equation} 
has frequently been used. Here, $\dot{\gamma}$, $D_g$, $p$, and $\rho_g$ are the shear strain rate, grain diameter, confining pressure, and bulk density of granular matter, respectively. The inertial number represents the ratio between shearing timescale $\dot{\gamma}^{-1}$ and grain rearrangement timescale due to the confining pressure $D_g/{\sqrt{p / \rho_g}}$. Thus, the shearing dominates the dynamics when $I$ is large while the confining pressure becomes essential in small $I$ regime. In a sense of ratio between shearing and confining, this form is more or less analogous to the Coulomb friction coefficient,
\begin{equation}
\mu =\frac{F}{N}, 
\label{eq:Coulomb_friction}
\end{equation}
where $\mu$, $F$, and $N$ are the friction coefficient, tangential force, and normal force applied to the sliding surface, respectively. Of course, $F$ and $N$ respectively correspond to shearing and confining in this form. Since $I$ involves the granular property $D_g$, it could be a relevant parameter to characterize $\mu$ for granular friction. 

For example, an empirical constitutive law using $I$ in dense granular flow regime has been proposed~\cite{Jop2006,Pouliquen2006}. According to these studies, the granular frictional coefficient can be written as, 
\begin{equation}
\mu= \mu_s +\frac{\mu_2 - \mu_s}{I_0/I+1},
\label{eq:Pouliquen_law}
\end{equation}
where $\mu_s$, $\mu_2$, and $I_0$ are empirical parameters corresponding to the friction coefficient at the limit $I\to 0$ and at $I\to\infty$, and the characteristic inertial number, respectively. Although this constitutive law is empirically obtained, it can be applied to various geometrical setups~\cite{Pouliquen2006}. 

Some other types of frictional constitutive law have also been proposed~\cite{daCruz2005,Hatano2007,Hatano2013,Kuwano2013}. Particularly, Kuwano et al. found a crossover from stress-weakening to stress-strengthening by varying $I$ in a very wide range~\cite{Kuwano2013}. Their form is written as, 
\begin{equation}
\mu = \mu_0 - \alpha \ln I + CI,
\label{eq:Kuwano_law}
\end{equation}
where $\mu_0$, $\alpha$, and $C$ are fitting parameters. While these constitutive laws (Eqs.~(\ref{eq:Pouliquen_law}) and (\ref{eq:Kuwano_law})) look different with each other, these values can be very similar when parameter values are appropriately adjusted in dense flow regime~\cite{Katsuragi2016}. Thus, the physical origin of these constitutive laws could be identical. To check the applicability and/or universality of the constitutive laws, experimental tests with various setups should be performed. Therefore, in this study, we are going to examine the granular constitutive law with a particular setup: very slow vertical withdrawing of a thin metal rod from a granular bed. 

This particular experimental setup has actually been conceived by considering the application of a very precise creep meter developed for biophysical measurement of plant cell wall rheology (improved version of the instrument developed in the previous study~\cite{Takahashi2006}). Using this very precise instrument, we perform a simple experiment and evaluate the geometry dependence of the granular frictional constitutive law.

\section{Experiment}
\begin{figure}[bt]
\centerline{\psfig{file=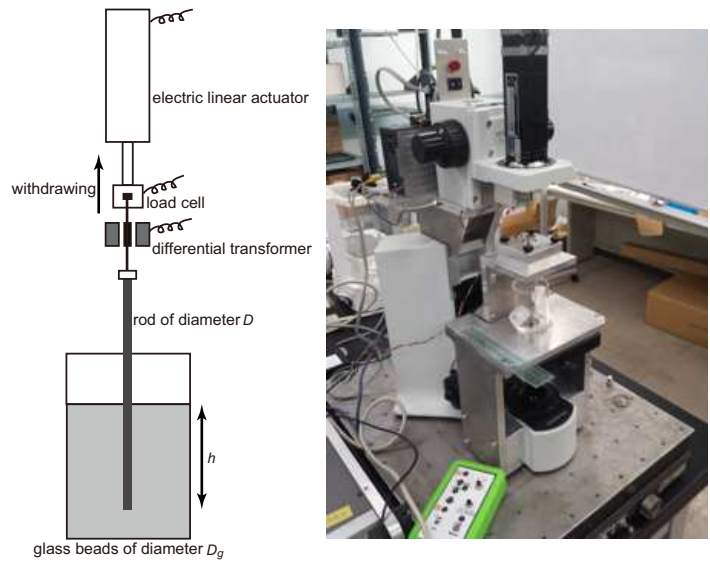,width=3.65in}}
\vspace*{8pt}
\caption{Schematic drawing (left) and picture (right) of the experimental apparatus. A rod is vertically withdrawn from a cylindrical container filled with glass beads. The force $F$ and displacement $z$ are precisely measured. In the right picture, the cylindrical container is empty.}
\label{fig:experiment_app}
\end{figure}

A schematic drawing and picture of the experimental apparatus are shown in Fig.~\ref{fig:experiment_app}. The system was constructed on the arm and base unit of a microscope system (OLYMPUS BX41). A stainless rod is vertically hung on a load cell (KYOWA ELECTRONIC INSTRUMENT LVS-5GA) to measure the pulling force. The vertical displacement of the rod $z$ is controlled by an electric linear ball screw actuator (NTN DMX3104SA-02+D 1-KY) and measured with a differential transformer (SEIYU ELECTRONIC DTD-3). The motion of the actuator is controlled with an in-house developed controlling system, and the position and pulling force of the rod are simultaneously recorded. Using this system called PCM-nano (nano-version Programmable Creep Meter), we can control the maximum pulling velocity $V_{\rm max}$ in the range of $V_{\rm max}=0.4$ - $16$ $\mu$m/s. Diameter of the rod $D$ is varied from $D=0.55$ to $2.0$ mm. To make a granular bed, we use roughly monodisperse glass beads of diameter $D_g=0.1$, $0.4$, or $0.8$ mm (AS-ONE, BZ01, BZ04, and BZ08). 

Glass beads are poured into a cylindrical container (inner diameter $46$ mm), in which the rod is placed before the pouring. Glass beads are slowly poured with care to keep the rod vertical. Height of the granular bed is always set $45$ mm, and the initial penetration depth of the rod in the granular bed is set $h_0=44$ mm. The packing fraction achieved by this procedure is $0.6$. 

\begin{figure}[bt]
\centerline{\psfig{file=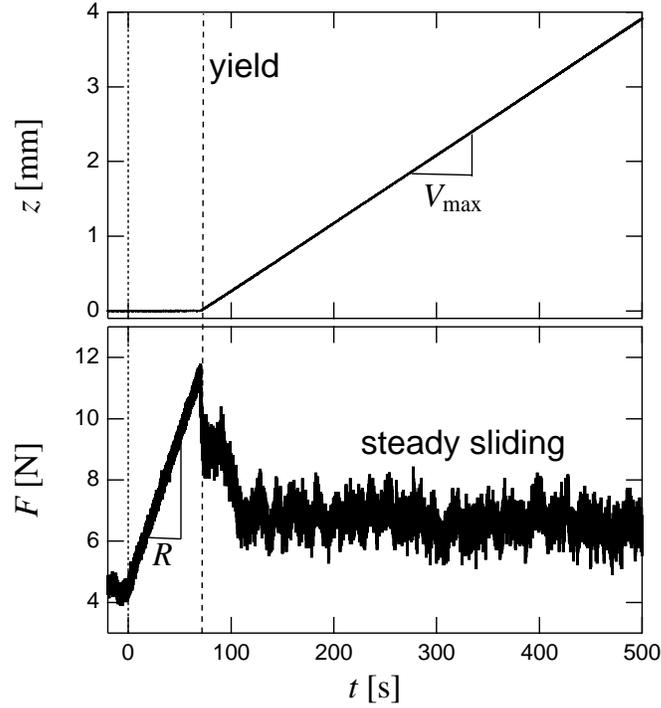,width=3.65in}}
\vspace*{8pt}
\caption{A typical example of raw data of $z(t)$ and $F(t)$. In the early state, the pulling force $F$ increases linearly with time $t$ obeying a control signal, $F=Rt$. In this regime, displacement $z$ is negligibly small. After the yielding, steady sliding can be observed. In this regime, $z$ linearly increases as $z=V_{\rm max}t$ and $F(t)$ is almost constant. Experimental conditions are $R=0.1$~mN/s, $V_{\rm max}=8$~$\mu$m/s, $D=1.2$~mm, and $D_g=0.4$~mm.}
\label{fig:example_data}
\end{figure}

Next, the experimental procedure of rod withdrawing is summarized. Right after the sample preparation, the rod is pulled. In this study, force-control condition is employed. Specifically, the increasing rate of pulling force $R$ is used as a main control parameter. $R$ is varied from $0.006$ to $6$ mN/s. The experimental protocol is as follows~(Fig.~\ref{fig:example_data}). From the initial time ($t=0$) defined by the beginning of pulling, the force is measured with a sampling rate ($100$ - $4000$ Hz). If the measured force $F$ is smaller than the control signal $Rt$, the rod is withdrawn by the linear actuator. The minimum withdrawing length set by the actuator is $4$ nm. When $F\geq Rt$, on the other hand, the rod is not withdrawn. During the experiment, force $F$ and vertical displacement of the rod $z$ are recorded on a PC through DAQ system (NI~9215). $z=0$ corresponds to the height level at $t=0$. Three experimental runs are carried out for each experimental condition to check the reproducibility.  

A typical example of the acquired data is shown in Fig.~\ref{fig:example_data}. In the early stage, static granular friction resists the pulling force. In this regime, $F$ increases as $F=Rt$, and the displacement $z$ is almost constant ($z\simeq 0$). When $Rt$ exceeds a certain limit, yielding of the granular bed occurs. Then, in this late stage, the sliding with a constant speed $V_{\rm max}$ can be observed. In this regime, granular dynamic friction balances with the pulling force. That is why the steady sliding with constant speed and force is clearly attained at $t>120$ s in Fig.~\ref{fig:example_data}. This steady sliding regime is particularly focused in this study since the analysis of steady sliding state is usually more understandable than that of static or transient state~\cite{Katsuragi2016}. This paper presents a first-step approach to the granular withdrawing friction in slow regime.

To characterize the steady sliding regime, we use friction coefficient $\mu$~(Eq.~(\ref{eq:Coulomb_friction})) and inertial number $I$~(Eq.~(\ref{eq:inertial_number})). Although $F$ is directly measurable, the normal force $N$ cannot be measured in this experiment. Here, we assume hydrostatic pressure for the origin of $N$. Then, $N$ is simply computed as
\begin{equation}
N= \int_0^h \rho_g g h' \pi D dh' = \frac{\pi \rho_g g D h^2}{2},
\label{eq:normal_force}
\end{equation} 
where $h=h_0-z$ is the contact length and $g=9.8$ m/s$^2$ is gravitational acceleration. Similar depth-proportional hydrostatic form for granular friction has been found in the impact drag force modeling~\cite{Katsuragi2007,Katsuragi2013b}, whereas the nonlinear drag force has also been reported in slow penetration experiments~\cite{Katsuragi2012a,Katsuragi2012b}. To compute the inertial number $I$, we have to estimate $\dot{\gamma}$ and $p$ as well. For $\dot{\gamma}$, we simply assume $V_{\rm max} \sim \dot{\gamma} D_g$, on the basis of previous experiment~\cite{Kuwano2013}. As for the pressure $p$, it depends on $z$ since we assume the hydrostatic form. For the sake of simplicity, here we use a representative value $p=\rho_g g h_0/\sqrt{2}$ which corresponds to the hydrostatic pressure at the depth of $h_0/\sqrt{2}$. Namely, we simply define the inertial number with measurable quantities as, 
\begin{equation}
I= \frac{V_{\rm max}}{\sqrt{p/\rho_g}} \sim \frac{V_{\rm max}}{\sqrt{gh_0}}.
\label{eq:I_redef}
\end{equation}
Note that, on the other hand, $N$ is a quadratic function of $h$. Using these assumptions and measured data, we can compute $\mu$ and $I$.

\section{Results and analyses}
\begin{figure}[bt]
\centerline{\psfig{file=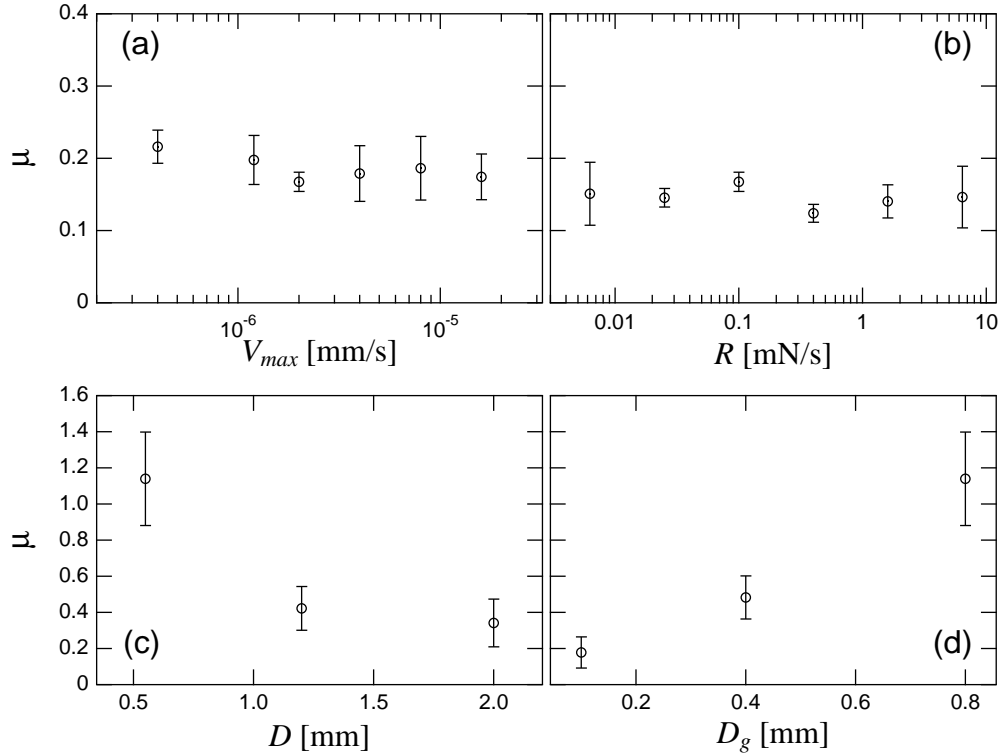}}
\vspace*{8pt}
\caption{Measured results of the friction coefficient $\mu$. $\mu$ is plotted as a function of (a)~$V_{\rm max}$, (b)~$R$, (c)~$D$, or (d)~$D_g$. One can confirm that $\mu$ is almost independent of $V_{\rm max}$ and $R$ while it strongly depends on $D$ and $D_g$. Experimental conditions are as following. For (a) and (b), $D$ and $D_g$ are fixed to $1.2$ and $0.4$ mm, respectively. $R=0.1$ mN/s and $V_{\rm max}=2$ $\mu$m/s are used for (a) and (b), respectively. For (c) and (d), $R$ and $V_{\rm max}$ are respectively fixed to $0.1$ mN/s and $2$ $\mu$m/s. $D_g=0.8$ mm and $D=0.55$ mm are employed for (c) and (d), respectively. Error bars (in all plots) indicate the standard deviation of three experimental runs.}
\label{fig:raw_data}
\end{figure}

In Fig.~\ref{fig:raw_data}, $\mu=F/N$ measured in the steady sliding regime is shown as a function of $V_{\rm max}$, $R$, $D$, or $D_g$. In order to compute the mean value of $\mu$, we take an average of fluctuating $\mu$ in the late steady regime for more than $10$ s. As can be seen in Fig.~\ref{fig:raw_data}(a,b), $\mu$ is almost independent of $V_{\rm max}$ and $R$. This means that the current experimental conditions correspond to the very slow (quasi-static) sliding regime in which the rate dependence of the frictional behavior cannot be observed even in steady (dynamic) frictional regime. However, $\mu$ depends on both $D$ and $D_g$~(Fig.~\ref{fig:raw_data}(c,d)). That is, the granular friction strongly depends on geometrical setup rather than the sliding rate. 

\begin{figure}[bt]
\centerline{\psfig{file=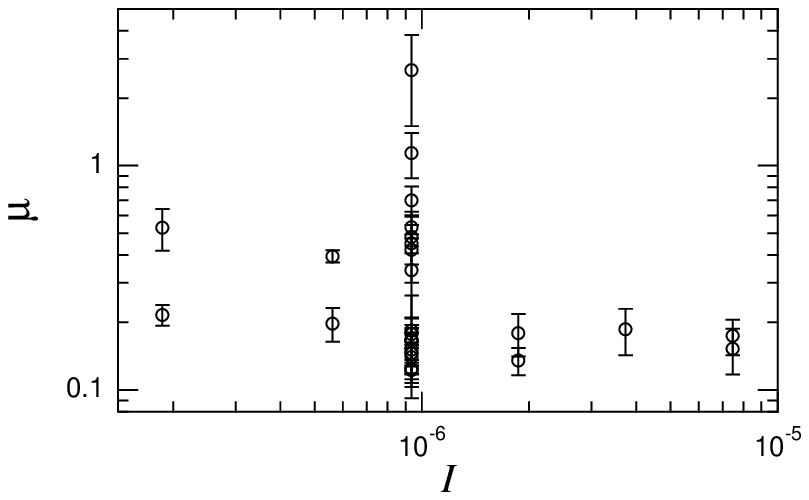,width=3.65in}}
\vspace*{8pt}
\caption{All the measured $\mu$ values are shown as a function of inertial number $I$. If $I$ is a relevant dimensionless parameter to describe the current phenomenon, the data should collapse to a master curve. However, the data considerably scatter even for the identical $I$ case. For instance, $\mu(I\simeq 10^{-6})$ looks a multiple-valued function.}
\label{fig:mu_I}
\end{figure}

This tendency can clearly be confirmed when we plot $\mu$ as a function of $I$~(Fig.~\ref{fig:mu_I}). If $I$ defined by Eq.(\ref{eq:I_redef}) is a relevant dimensionless number to describe the granular frictional constitutive law even in slow sliding regime, all the data should collapse onto a single master curve. However, as can be confirmed in Fig.~\ref{fig:mu_I}, the data considerably scatter and cannot be expressed by a single-valued function. Therefore, we have to seek another relevant dimensionless variable for slow granular friction. This strong geometry dependence is contrastive to the traditional granular frictional constitutive laws~\cite{GDRMiDi2004} or rate- and state-dependent constitutive laws~\cite{Marone1998}. 

\begin{figure}[bt]
\centerline{\psfig{file=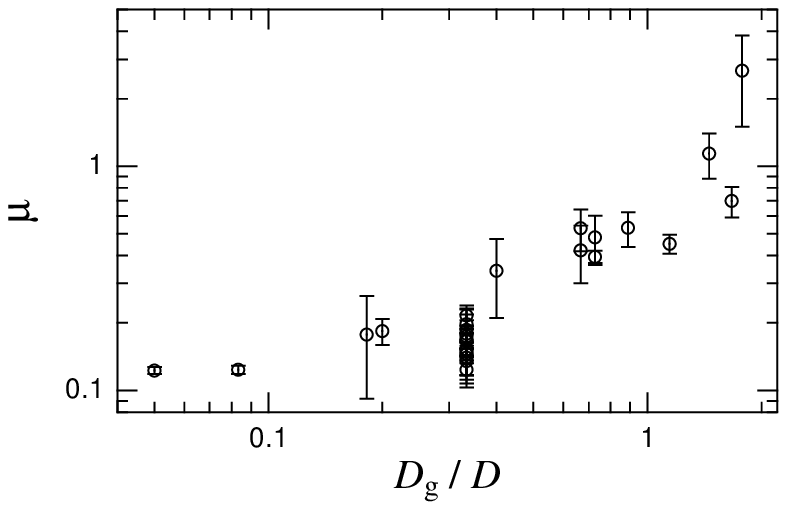,width=3.65in}}
\vspace*{8pt}
\caption{All the measured $\mu$ data are plotted by a diameter ratio $D_g/D$. Since $\mu$ strongly depends on $D$ and $D_g$, this plot seems to better explain the data behavior than Fig.~\ref{fig:mu_I}. In addition, $\mu$ exceeds unity in the large grain regime. In such a large $D_g/D$ regime, the assumption of hydrostatic pressure could be inappropriate.}
\label{fig:mu_DgD}
\end{figure}

The geometry dependence can actually be unified by a simple dimensionless number. In Fig.~\ref{fig:raw_data}(c,d), $D$ and $D_g$ dependences of $\mu$ clearly show opposite tendency. Thus, $D_g/D$ could be a relevant dimensionless number. In Fig.~\ref{fig:mu_DgD}, the relation between $\mu$ and $D_g/D$ is plotted. As expected, one can confirm a data collapse to a master constitutive relation. Therefore, we can conclude that the geometrical condition is much more crucial to estimate the granular friction at the slow sliding limit, at least in the current experimental setup: a thin rod vertically withdrawn from a granular bed. 

\section{Discussion}
Let us consider the physical meaning of the experimentally obtained constitutive law shown in Fig.~\ref{fig:mu_DgD}. First, $\mu$ is an increasing function of $D_g/D$. At the small $D_g/D$ limit, $\mu$ seems to approach an asymptotic value around $\mu \simeq 0.1$. This could correspond to the limit of continuum approximation due to the small grain size. In contrast, $\mu$ tends to increase significantly when $D_g/D$ is large. Particularly, $\mu$ exceeds unity when $D_g$ is greater than $D$. This estimate is intuitively incorrect since we consider the cohesionless glass beads and stainless rod. The reason for this unexpectedly large $\mu$ value is most likely the inappropriate estimate of confining pressure $p$. Although we have assumed a hydrostatic pressure for $p$, this assumption is not always satisfied in granular friction. In general, the internal stress is randomly scattered in granular matter. This stress scattering can directly be observed by photoelastic disk systems~(e.g.~\cite{Iikawa2015,Iikawa2016}). In bulk granular matter, this stress scattering could cause the wall support called Janssen effect~\cite{Janssen1895,Sperl2006}, which results in the pressure saturation in a deep part of granular column. Although the current experimental setup is wide enough to neglect the Janssen effect (the depth and diameter of the container are comparable), the stress scattering might cause different type of peculiar behavior in granular friction.

\begin{figure}[bt]
\centerline{\psfig{file=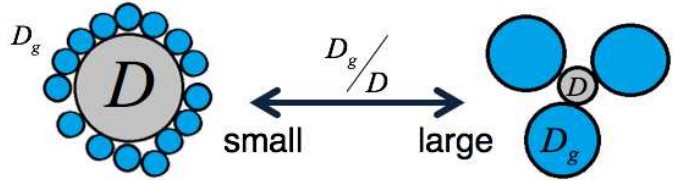,width=3.65in}}
\vspace*{8pt}
\caption{Qualitative images of the cross-section. In the small $D_g/D$ regime, a lot of grain contacts and surrounding grains would result in the continuum-like situation. However, when $D_g/D$ is large, there are very few grains contacting with the rod. As a result, granular peculiarity would be enhanced in this regime. We consider that the shear band structure is formed and it affects the frictional property in this regime.
}
\label{fig:DgD_img}
\end{figure}

Considering the stress scattering within a granular bed, we can qualitatively understand the peculiar behavior of the experimentally obtained constitutive law. In Fig.~\ref{fig:DgD_img}, cross-sectional schematics of contacts between glass beads and rod are presented. When $D_g/D$ is small, a lot of grains contact with the rod~(Fig.~\ref{fig:DgD_img}~left). In such situation, randomly scattered stresses are averaged out in the vicinity of the rod and a hydrostatic approximation would provide a reasonable estimate. However, if $D_g/D$ is large, the number of grains contacting with the rod becomes very small~(Fig.~\ref{fig:DgD_img}~right). As a consequence, peculiar nature of granular stress scattering is significantly enhanced due to the exaggerated discreteness of granular matter.

\begin{figure}[bt]
\centerline{\psfig{file=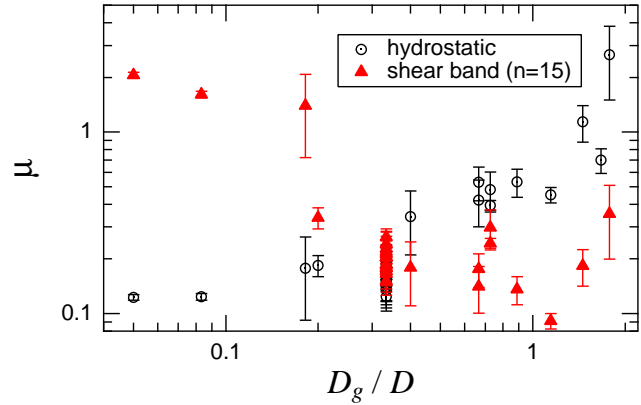,width=3.65in}}
\vspace*{8pt}
\caption{Comparison of hydrostatic-pressure-based $\mu$ (black circles) and shear-band-based $\mu$ with $n=15$ (red triangles). In the small $D_g/D~(<0.4)$ regime, $\mu$ is stable around $0.1$ by the hydrostatic model. In the large $D_g/D~(>0.4)$ regime, on the other hand, $\mu$ fluctuates around $0.1$ - $0.2$ by the shear band model. They show a crossover around $D_g/D\simeq 0.3$.
}
\label{fig:improved_mu}
\end{figure}

This granular peculiarity could result in the shear band structure at large $D_g/D$. By assuming a simple annular shear band for a source of normal force applying to the rod, we can improve the estimate of $\mu$. Here we introduce a thickness of shear band $n$ in grain-diameter unit, i.e., actual shear band thickness is $nD_g$. And, we assume that the normal force acting to the rod is proportional to mass of this shear band structure. Then, the modified $\mu$ is written as,
\begin{equation}
\mu = \frac{2F}{\pi \rho_g g h_0 \left[ \left(nD_g +D/2\right)^2 - \left(D/2\right)^2 \right]}.
\label{eq:modified_nu}
\end{equation}  
By varying $n$, we find that $n\simeq 15$ results in $\mu\simeq 0.1$ even in large $D_g/D$ regime (Fig.~\ref{fig:improved_mu}). Here, we use an approximation $h\simeq h_0$. Actually, this shear-band-thickness value ($n\simeq 15$) is close to that observed in granular heap flow~\cite{Katsuragi2010b}. Although the reason of this coincidence of the shear band thickness is not clarified, it provides a presumptive evidence for the validity of the current modeling. The shear band structure could also be induced even in small $D_g/D$ regime. However, the shear band model yields too large $\mu$ in small $D_g/D$ regime~(Fig.~\ref{fig:improved_mu}). This suggests that the hydrostatic pressure or combination of hydrostatic pressure and shear band is necessary for explaining the behavior in small $D_g/D$ regime. 

After considering the geometry dependence through $D_g/D$, the rate dependence can probably be quantified by the inertial number. The combination of $D_g/D$ and $I$ (e.g., $ID/D_g$) or the redefinition of $I$ might be able to provide a possible relevant dimensionless number. Systematic experiment with a wider rate variation is a necessary next step to reveal the universal granular frictional constitutive law.

In this experiment, we only analyze the steady sliding regime to characterize the dynamic (but slowly sliding) frictional property for granular matter. However, we can also characterize the static friction by identifying the peak of force curve typically shown in Fig.~\ref{fig:raw_data}. The study of static friction is an important future work. 
Furthermore, in some experiments, we observe some precursive slipping events at the early stage of withdrawing. The statistical analysis of precursive events will be an interesting research topic too. Moreover, the study of geometry effect by varying the shape of withdrawn object might also be crucial to understand the nature of granular friction. 

\section{Conclusion}
In this study, granular frictional constitutive law in very slow (quasi-static) but steady sliding regime was experimentally investigated. Using a very precise creep meter, a thin rod was vertically withdrawn from a glass beads bed. The pulling force $F$ was controlled to be gradually increased, $F=Rt$. At a certain force level of $F$, the yielding followed by steady sliding is induced. In the steady sliding regime, we measure the friction coefficient $\mu$. The measured $\mu$ cannot be a function of a simply defined inertial number $I$ which has been used for characterizing granular friction in many previous works. Instead, we found that the ratio between grain diameter $D_g$ and withdrawn rod diameter $D$ is a relevant parameter to understand the experimentally obtained $\mu$. In addition, $\mu$ can be greater than unity if we simply assume granular hydrostatic pressure for the source of normal force. This seemingly counterintuitive property can be understood by considering the shear band structure in large $D_g/D$ regime. In small $D_g/D$ regime, on the other hand, hydrostatic pressure approximation seems to work well. This means that the effective contact number $D_g/D$ is an indicator of granular discreteness in the current experimental setup~(a thin rod withdrawing case).

\section*{Acknowledgements}

The authors would like to thank S.~Watanabe, H.~Kumagai, S.~Sirono, and T.~Morota for helpful discussion and comments. They are grateful to T.~Torii of his helpful advice in the planning of PCM-nano. They thank Dr.~K.~Takahashi for his help in an examination of performance of the PCM-nano.

\section*{References}

\bibliography{fgf}

\begin{thebibliography}{10}

\bibitem{daCruz2005}
Fr\'ed\'eric da~Cruz, Sacha Emam, Micha\"el Prochnow, Jean-No\"el Roux, and
  Fran\ifmmode \mbox{\c{c}}\else~\c{c}\fi{}ois Chevoir.
\newblock Rheophysics of dense granular materials: Discrete simulation of plane
  shear flows.
\newblock {\em Phys. Rev. E}, 72:021309, 2005.

\bibitem{GDRMiDi2004}
{GDR~MiDi}.
\newblock On dense granular flows.
\newblock {\em Eur. Phys. J. E}, 14:341--365, 2004.

\bibitem{Hatano2007}
Takahiro Hatano.
\newblock Power-law friction in closely packed granular materials.
\newblock {\em Phys. Rev. E}, 75:060301, Jun 2007.

\bibitem{Hatano2013}
Takahiro Hatano and Osamu Kuwano.
\newblock Origin of the velocity-strengthening nature of granular friction.
\newblock {\em Pure Appl. Geophys.}, 170:3--11, 2013.

\bibitem{Iikawa2015}
N.~Iikawa, M.~M. Bandi, and H.~Katsuragi.
\newblock Structural evolution of a granular pack under manual tapping.
\newblock {\em J. Phys. Soc. Jpn.}, 84:094401, 2015.

\bibitem{Iikawa2016}
N.~Iikawa, M.~M. Bandi, and H.~Katsuragi.
\newblock Sensitivity of granular force chain orientation to disorder-induced
  metastable relaxation.
\newblock {\em Phys. Rev. Lett.}, 116:128001, 2016.

\bibitem{Janssen1895}
H.~A. Janssen.
\newblock Versuche {\"u}ber getreidedtruck in silozellen.
\newblock {\em Z. Ver. Dtsh. Ing.}, 39:1045--1049, 1895.

\bibitem{Jop2006}
P.~Jop, Y.~Forterre, and O.~Pouliquen.
\newblock A constitutive law for dense granular flows.
\newblock {\em Nature}, 441:727--730, 2006.

\bibitem{Katsuragi2010b}
H.~Katsuragi, A.~R. Abate, and D.~J. Durian.
\newblock Jamming and growth of dynamical heterogeneities versus depth for
  granular heap flow.
\newblock {\em Soft Matter}, 6:3023--3029, 2010.

\bibitem{Katsuragi2007}
H.~Katsuragi and D.~J. Durian.
\newblock Unified force law for granular impact cratering.
\newblock {\em Nat. Phys.}, 3:420--423, 2007.

\bibitem{Katsuragi2012b}
Hiroaki Katsuragi.
\newblock Material, preparation, and cycle dependence of pressure behavior in a
  slowly plunged granular column.
\newblock {\em Chem. Eng. Sci.}, 76:165--172, 2012.

\bibitem{Katsuragi2012a}
Hiroaki Katsuragi.
\newblock Nonlinear wall pressure of a plunged granular column.
\newblock {\em Phys. Rev. E}, 85:021301, Feb 2012.

\bibitem{Katsuragi2016}
Hiroaki Katsuragi.
\newblock {\em Physics of Soft Impact and Cratering}.
\newblock Springer, 2016.

\bibitem{Katsuragi2013b}
Hiroaki Katsuragi and Douglas~J. Durian.
\newblock Drag force scaling for penetration into granular media.
\newblock {\em Phys. Rev. E}, 87:052208, May 2013.

\bibitem{Kuwano2013}
Osamu Kuwano, Ryosuke Ando, and Takahiro Hatano.
\newblock Crossover from negative to positive shear rate dependence in granular
  friction.
\newblock {\em Geophys. Res. Lett.}, 40:1295--1299, 2013.

\bibitem{Marone1998}
Chris Marone.
\newblock Laboratory-derived friction laws and their application to seismic
  faulting.
\newblock {\em Annu. Rev. Earth Planet. Sci.}, 26:643--696, 1998.

\bibitem{Pouliquen2006}
O.~Pouliquen, C.~Cassar, P.~Jop, Y.~Forterre, and M.~Nicolas.
\newblock Flow of dense granular material: towards simple constitutive laws.
\newblock {\em J. Stat. Mech.}, page P07020, 2006.

\bibitem{Sperl2006}
Matthias Sperl.
\newblock Experiments on corn pressure in silo cells – translation and
  comment of janssen's paper from 1895.
\newblock {\em Granular Matter}, 8:59--65, 2006.

\bibitem{Takahashi2006}
Koji Takahashi, Shinya Hirata, Nobuo Kido, and Kiyoshi Katou.
\newblock Wall-yielding properties of cell walls from elongating cucumber
  hypocotyls in relation to the action of expansin.
\newblock {\em Plant and Cell Physiology}, 47(11):1520--1529, 2006.

\end{thebibliography}

\end{document}